# Integrating spectral and morphological plant features with decision-tree models for early-season cotton biomass and nitrogen status estimation from multi-year UAV data

**Vaishali Swaminathan[a,b,*], Nithya Rajan[c], J Alex Thomasson[a,d], Amrit Shrestha[a,e], Karem Meza Capcha[a,f], Robert Hardin[a], Pramod Pokhrel[c,g]**

[a] Department of Biological and Agricultural Engineering, Texas A&M University, College Station, Texas, USA.

[b] Utah Water Research Laboratory, Utah State University, Logan, Utah, USA.

[c] Department of Soil and Crop Sciences, Texas A&M University, College Station, Texas, USA.

[d] Mississippi Water Resources Research Institute, Mississippi State University, Starkville, Mississippi, USA

[e] Department of Agricultural and Biological Engineering, Mississippi State University, Mississippi State, Mississippi, USA

[f] Department of Viticulture and Enology, University of California, Davis,California, USA

[g] Northwest Research- Extension Center, Department of Agronomy, Kansas State University, Colby, KS, USA

* Corresponding author: Vaishali Swaminathan (vaishali.swaminathan@usu.edu)

## Abstract

Precision nitrogen (N) management (PNM) for cotton (*Gossypium hirsutum*) requires in-season monitoring of crop growth parameters and N status indicators to decide fertilizer timing, placement, and application rates for optimal canopy development and yield. This study developed remote sensing and machine learning-based methods to estimate cotton dry biomass weight (DBW), plant N uptake (PNU), plant N concentration (PNC), critical N dilution ($N_c$), and nitrogen nutrition index (NNI) to support PNM. To achieve this, a three-year field-based N-management study was conducted and unmanned aerial vehicle (UAV)-based multispectral images were acquired between early vegetative growth and flowering stages, critical for fertilizer applications. Spatiotemporally consistent spectral and morphological plant features, including plant height (PH) and fractional canopy cover (FCC), provided reliable model training inputs. DBW, PNU, and PNC estimates from simple regression using vegetation indices (VIs), multiple linear regression (MLR) combining VIs, PH, and FCC, and decision-tree models – random forest regression (RFR) and extreme gradient boosting (XGB) – combining spectral reflectance, PH, and FCC were evaluated using trial-held-out (THO) and leave-one-year-out (LOYO) validation methods. The best validation accuracies were from $RFR_{THO}$ ($R^2$ = 0.88 and MAPE = 23.14% for DBW; $R^2$ = 0.84 and MAPE = 20.61% for PNU; $R^2$ = 0.85 and MAPE = 7.82% for PNC) and $XGB_{THO}$ ($R^2$ = 0.87 and MAPE = 21.91% for DBW; $R^2$ = 0.81 and MAPE = 21.40% for PNU; $R^2$ = 0.86 and MAPE = 7.66% for PNC). The decision-tree models also improved LOYO-validation of seasonally invariable parameters like DBW and PNU. Both RFR and XGB trained on parsimonious subsets of spectral and morphological features, showing that combining canopy morphology with spectral health indicators improved N status estimation. $N_c$ was calculated from model estimated DBW and PNC for high-yielding, medium-to-tall cotton varieties grown in the Texas Coastal Plains and validated using ground-truth biomass measurements. NNI derived from $XGB_{THO}$ outputs performed marginally better than NNI from $RFR_{THO}$ in identifying N-deficient plots ($F1_{RFR}$ = 0.70; $F1_{XGB}$ = 0.75) and multi-level N-stress categorization ($F1_{RFR}$ = 0.52; $F1_{XGB}$ = 0.59). This study provides a framework for estimating crop growth parameters from standardized, well-calibrated UAV-multispectral images, which can be used for cultivar- and region-specific $N_c$ and NNI estimation and fertilizer decision-support.

# 1. Introduction

Nitrogen (N) fertilizer is often used excessively in agricultural production systems because the perceived economic benefits from yield increase are considered greater than the additional fertilizer and environmental costs associated with it. Excessive fertilizer application and the simultaneous underutilization of nutrients by crops cause environmental pollution through pathways like denitrification loss, ground-water leaching, and nutrient runoffs into waterbodies (Schulte-Uebbing et al. 2022). The increase in fertilizer consumption can results in a 60% increase in environmental pollution from agricultural sources by 2050 (Martínez-Dalmau et al. 2021). Hence, sustainable agricultural production strategies like precision N management (PNM) emphasize the application of right source and amount of fertilizer, at the right place and time to ensure efficient N uptake by crops while minimizing loss of N to the environment.

Studies have shown that about 30% of the total N required for cotton (*Gossypium hirsutum*) production is consumed between seedling emergence and first bloom, and the remaining 70% is used before peak bloom (Flis 2019). During this period, active translocation of N from older leaves to newer leaves and fruiting organs increases metabolic N demand to sustain vegetative growth. Therefore, N stress accelerates the breakdown of leaf chlorophyll (chlorosis), causing early onset of senescence, stunted canopies, and yield loss. Conversely, excessive N affects cotton production by causing abundant vegetative growth, delayed maturity, and profit decline from missed ideal harvest periods (Rochester and Constable 2020; Sui et al. 2017). Hence, identifying optimal N requirements during early vegetative growth, squaring, and flowering stages when N uptake increases is critical for adapting PNM strategies for cotton.

Unmanned aerial vehicles (UAV) remote sensing is widely used for PNM in large agricultural fields with soil and terrain variabilities, because high resolution imaging sensors (multispectral and hyperspectral) can efficiently monitor spatial and temporal patterns in crop growth and N requirements. Canopy spectral signatures provide insights into leaf pigment, cellular structure, turgidity, and canopy architecture, which are influenced by plant N uptake. Specifically, N availability impacts chlorophyll (*Chl-a* and *Chl-b*) absorption in the blue (450 nm) and red (670 nm) wavelengths and leaf cellular structure that determines near-infrared (NIR) (760 to 1300 nm) reflectance (Wong 2023). By combining targeted wavebands, spectral vegetation indices (VIs) provide valuable insights into canopy physiology and structural traits, thus also serving as indicators of plant N status, N uptake, and total biomass accumulation (Colorado et al. 2020; Li, Fei et al. 2014). Furthermore, photogrammetry and computer-vision algorithms like structure from motion, simultaneous localization and mapping, and multi-view stereo estimate morphological features like plant height and 3D canopy architecture with high accuracy (Jiang, S. et al. 2020; Li, Zhengkun et al. 2025; Wu et al. 2022). Since N availability affects both physiological and structural plant properties, fusing different modalities, like spectral data with textural features (Li, Minghua et al. 2024), morphological plant features (Li, Jiating et al. 2018; Lu et al. 2021; Zheng et al. 2022), and management and environmental data (Jiang, J. et al. 2022; Ma et al. 2026) broadens feature diversity and improves overall crop biomass and N status estimation accuracy.

Machine learning models trained with multiple remotely sensed features are widely used for non-destructive, in-season agricultural parameter estimation. Several feature selection or reduction methods, such as the principal component analysis (PCA), Pearson correlation, recursive feature elimination, ElasticNet, and Boruta- shapely additive explanation (SHAP) help avoid model overfitting. Li et al. (2025) reported improvements in cotton N concentration estimated by random forest regression (RFR) models trained on a reduced set of spectral features obtained hierarchically with ElasticNet and Boruta-SHAP. Among various statistical and machine learning models including linear regression, principal component regression, gaussian process regression, support vector machine, and artificial neural networks, decision tree models like RFR, gradient boosting regression, and extreme gradient boosting regression (XGB) consistently performed well in estimating plant growth parameters like biomass (Dhakal et al. 2023; Zheng et al. 2022; Zhuo et al. 2024) and plant or leaf N content (Chen, X. et al. 2023; Cui et al. 2025; Peng et al. 2024; Tian et al. 2024). However, most crop N status models relied entirely on multicollinear spectral features that represented similar physiological attributes. This is further compounded by the sensitivity of VIs to cultivars differences (Avola et al. 2019; Hatfield and Prueger 2010; Hazratkulova et al. 2012), which restricts generalizability and transferability of VI-based models. Karaca et al. (2025) also showed that VI-based leaf N estimation was influenced by crop type, cultivar, and phenological stage, indicating that development of generalizable models would require datasets representing substantial physiological and environmental variability.

For variable-rate fertilizer decisions, which are crucial for real-time PNM, relative metrics like sufficiency index compare VIs from target areas to VIs from non-limiting reference strips to improve nitrogen use efficiency (Martins et al. 2020; Stamatiadis et al. 2020). The nitrogen nutrition index (NNI), a more generalized index independent of in-field reference strips, compares actual N concentration in plant tissues to the critical N ($N_c$) required for optimum biomass accumulation (Ciampitti et al. 2022). However, NNI requires established $N_c$ dilution curves, which to the best of our knowledge, was not sufficiently explored in recent years for cotton cultivars grown in USA. A few studies from China (Hou et al. 2021; Pei et al. 2023; Qin et al. 2025; Wang et al. 2025) showed considerable variations in coefficients describing the dilution curves, indicating possible genotype and environment effects on $N_c$ and NNI. While decision-trees are generally good at estimating regional or variety-specific NNI for cotton (Jia et al. 2025; Pei et al. 2023) and other crops (Jiang, J. et al. 2022; Qiu et al. 2021; Zhang et al. 2025), this study modeled fundamental biophysical crop parameters from which variety-specific $N_c$ dilution and NNI can be estimated.

This study investigated the integration of spectral and morphological plant features derived from UAV-borne multispectral images to estimate dry biomass weight (DBW), plant N uptake (PNU), plant N concentration (PNC), and NNI during critical early (vegetative to flowering) cotton growth stages, for timely fertilizer management. The primary objectives of this study included (1) estimating $N_c$ dilution for high-yielding, medium-to-tall cotton varieties grown in the Texas Coastal Plains, (2) developing geospatial image processing workflows to extract physical features like plant count, individual plant height (PH), and fractional canopy cover (FCC) from aerial remote sensing products, (3) developing physically interpretable decision-tree models to estimate crop growth and N status from important spectral and morphological plant features, and (4) evaluating effectiveness of NNI as a fertilizer management decision support tool.

## 2. Materials and Methods

### 2.1 Field Experiments

A three-year cotton N management study was conducted during the summers of 2020, 2021, and 2022 at Texas A&M University's Agrilife Research Farm near College Station, Texas, USA (Figure 1). N recommendations for the high-yielding, medium-to-tall cotton varieties – PHY 350 W3FE in 2020 and 2021; PHY 332 W3FE in 2022 – was 122 kg N/ha (Bronson 2016). Since, soil at the site had low baseline N concentration (0 to 9 ppm), urea ammonium nitrate (UAN) fertilizer was incorporated into the soil at four N rates (kg N/ha) – 0 (Control), 56 (low), 112 (recommended), 168 (high) – split at emergence and squaring (Table 1). Treatments were assigned following randomized block design with three replicates in each growing season. While the 2021 and 2022 experiments had all six treatments listed in Table 1, only four treatments (T1, T2, T5, and T6) were administered in the initial study during 2020. The plots were irrigated to meet 90% crop evapotranspiration demands, and pest and weed management were implemented to prevent compounding effects of other biotic and abiotic stresses. However, uncontrollable factors like variations in rainfall and temperatures impacted seasonal plant growth patterns and yield outcomes (Table 2).

**Table 1 The split nitrogen (N) application rates (kg/ha) used in the cotton N management studies during 2020, 2021, and 2022.**

| Treatments | Split Application Rates (initial +side dress) (kg/ha) | Treatment Classification | Treatment Years |
|---|---|---|---|
| T1 | 0 + 0 | control | All |
| T2 | 0 + 56 | low | All |
| T3 | 0 + 112 | recommended | 2021, 2022 |
| T4 | 56 + 0 | low | 2021, 2022 |
| T5 | 112 + 0 | recommended | All |
| T6 | 56 + 112 | high | All |

**Table 2 Field conditions including average residual soil nitrogen content (ppm), cumulative rainfall (mm), and minimum and maximum temperature (°C) observed during the 2020, 2021, and 2022 cotton nitrogen management field trials. The average seed cotton yield (kg/ha) and the harvest dates reported in days after planting (DAP) indicate the variability in outcomes of the controlled nitrogen management experiment due to external environmental factors.**

| Year | Residual soil N Avg (ppm) | Cumulative Rainfall (m) | Temperature (°C) | | Harvest (DAP) | Seed Cotton Yield Avg (kg/ha) |
|---|---|---|---|---|---|---|
| | | | Min | Max | | |
| 2020 | 7.00 | 0.36 | 9.83 | 40.11 | 168 | 1241.64 |
| 2021 | 5.10 | 0.64 | 8.20 | 37.40 | 169 | 3539.30 |
| 2022 | 2.33 | 0.00 | 8.20 | 41.84 | 129 | 1007.98 |

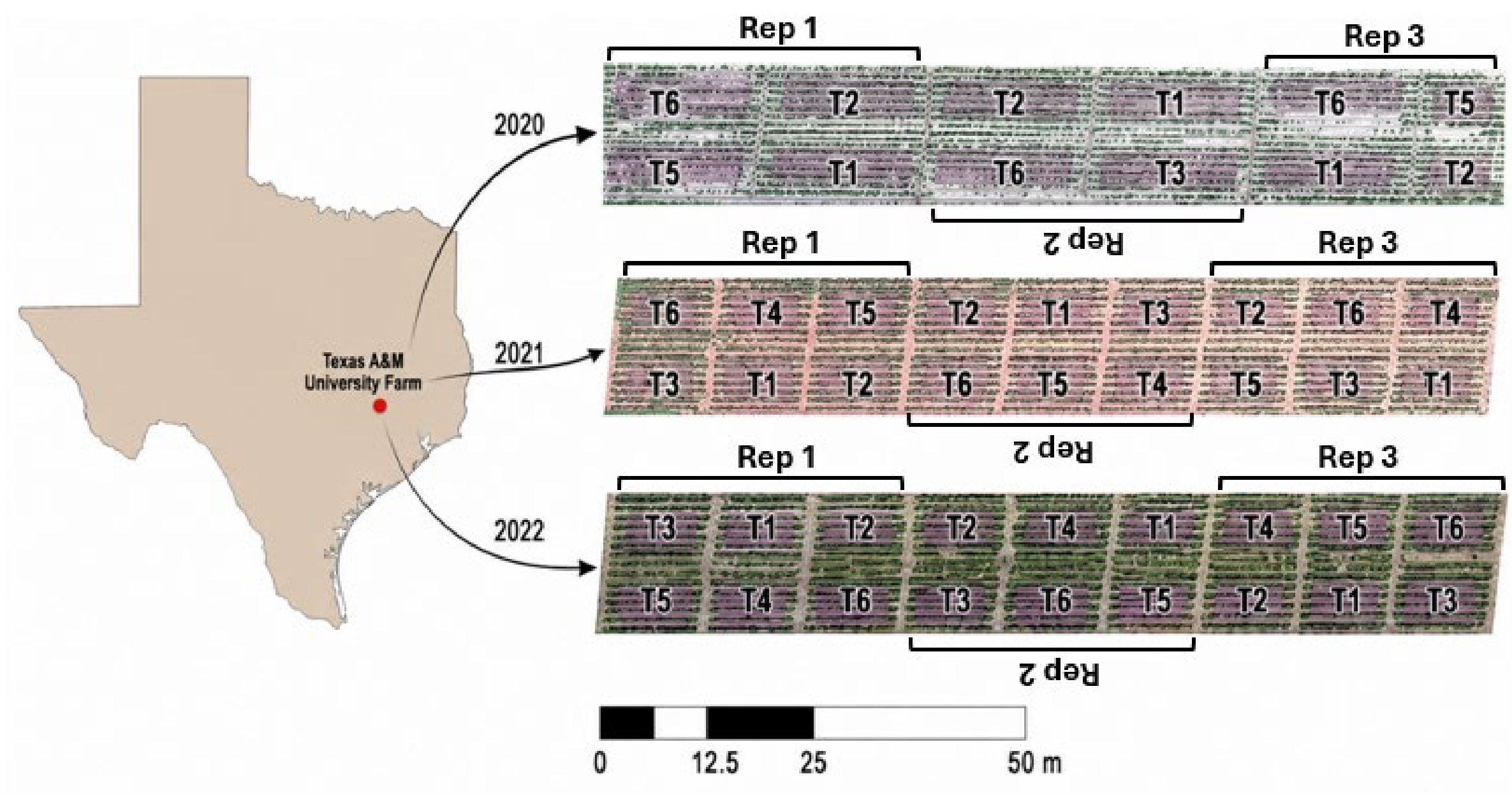

**Figure 1 Study location and nitrogen (N) management trial designs for 2020, 2021, and 2022.**

## 2.2 Data collection and ground-truthing

Remote sensing and ground-truth data were collected once at emergence and at 6 – 10 days intervals during early vegetative (St0, St1, and St2), squaring (St3 and St4), and flowering (St5) growth stages. A Micasense Red-Edge-3 camera (EagleNXT, Wichita, Kansas, USA), integrated with a DLS2 (downwelling light sensor + GPS/IMU) device, was mounted on a Matrice 100 UAV (DJI, Shenzhen, Guangdong, China) to acquire high-resolution (2.08 cm/pixel) multispectral images, within ± 1.5 hours of local solar noon. All images had 80% forward and 70% lateral overlaps, with the exceptions of 2020 flights having only 60% lateral overlap. Fixed exposure time and gain settings were used instead of autoexposure to prevent pixel saturation and ensure spatiotemporal radiometric accuracy (Swaminathan et al. 2024a). In-field calibration panels consisting of black (8%), gray (~25%), and white (2020 & 2021: ~50%; 2022:85%) reflectance targets were used for atmospheric correction and radiometric calibration. An RTK GNSS receiver (Reach RS2, Emlid, Budapest, Hungary), with < 0.02 m precision, recorded geographic (XYZ) coordinates of ground control points placed across the field for georectification and elevated control points (93 cm above ground) placed on raised row beds within trials for plant height calibration.

Above-ground plant height (PH, m) was measured for fifteen randomly selected plants from each plot. Plant stand count ($n_{plant}$) was obtained from five random 1-$m^2$ zones within each plot during emergence, from which average plant population density (PPD, plants/ $m^2$) was determined (2020: 4.83 plants/$m^2$; 2021: 8.00 plants/$m^2$; 2022: 7.55 plants/$m^2$). Also, total plot-level plant count was manually obtained from the 2020 aerial images for further validation of plant count. Biomass organs (leaves, stems, and flowers/fruits), obtained from six plants per plot per data collection event, were separated and dried until constant weight. Total dry biomass weight (DBW, g/plant) was calculated by adding individual organ weights ($w_{organ}$ g/plant) (equation 1). The dried samples were pulverized and analyzed for N concentration ($N_{organ}$ %) through combustion (McGeehan and Naylor 1988; Nelson and Sommers 1973), from which total plant N uptake (PNU, g/plant) (equation 2) and plant N concentration (PNC, %) (equation 3) were calculated. DBW and PNU were converted from g/plant to g/$m^2$ using PPD measurements.

$$DBW = w_{Leaf} + w_{Stem} + w_{Flower} \tag{1}$$

$$PNU = N_{Leaf} * w_{Leaf} + N_{Stem} * w_{Stem} + N_{Flower} * w_{Flower} \tag{2}$$

$$PNC = \frac{PNU}{DBW} * 100 \tag{3}$$

$N_c$ was interpolated by fitting a power regression equation between PNC (%) and DBW (Mg/ha) (Ciampitti et al. 2022), which served as reference for NNI (equation 4). Usually, NNI < 1 indicates N stress, NNI = 1 indicates N sufficiency, and NNI > 1indicates excessive N availability in plant tissues.

$$NNI = \frac{PNC}{N_c} \tag{4}$$

**Table 3 Vegetation indices (VIs) formulation and their specific applications as listed in Bajocco et al. (2022). ρ represents the calibrated reflectance in the respective bands.**

| Index | Formula | Applications |
|---|---|---|
| Normalized difference vegetation index (NDVI) | $\frac{\rho_{NIR} - \rho_{red}}{\rho_{NIR} + \rho_{red}}$ | Canopy structure, biomass |
| Normalized difference red edge index (NDRE) | $\frac{\rho_{NIR} - \rho_{rededge}}{\rho_{NIR} + \rho_{rededge}}$ | Chlorophyll, nitrogen, canopy structure |
| Green Normalized difference Vegetation Index (GNDVI) | $\frac{\rho_{NIR} - \rho_{green}}{\rho_{NIR} + \rho_{green}}$ | Chlorophyll |
| Chlorophyll index red edge ($CI_{rededge}$) | $\frac{\rho_{NIR}}{\rho_{rededge}} - 1$ | Chlorophyll and nitrogen |
| Excess green minus red index (ExGR) | $3\rho_{green} - 2.4\rho_{red} - \rho_{blue}$ | Distinguish canopy and soil |

2.3 Multispectral image processing and feature extraction workflows

Multispectral images affected by cloud cover changes were preprocessed using methods described by Swaminathan et al. (2024a), where in-flight radiometric correction and reflectance calibration were performed using DLS2 recorded instantaneous irradiance measurements and in-field reflectance calibration targets, respectively. Multispectral orthomosaics and digital elevation maps (DEMs) photogrammetrically processed from Metashape Pro (Agisoft LLC, St. Petersburg, Russia). A workflow consisting of computer vision and geospatial tools extracted (1) canopy masks and FCC, (2) VIs and band-wise spectral reflectance (ρ), and (3) plant count, PPD and PH. Plant canopy masks were obtained by performing adaptive thresholding on excess-green-minus-red (ExGR) maps. VIs listed in Table 3 that are sensitive to canopy structure, leaf nitrogen content, and chlorophyll concentration were averaged at plot-level. $VI_{mixed}$ containing both plant and soil pixels and $VI_{canopy}$ containing only canopy pixels were derived to assess if soil background affected the association between VIs and specific plant characteristics. Plant canopy area was calculated as product of pixel area ($2.08 \times 2.08 \times 10^{-4}$ $m^2$) and canopy pixel count, and FCC was obtained as the ratio of canopy area and total plot area. Canopy masks obtained at emergence were morphologically transformed (dilation and skeletonization) and vectorized to identify individual plant centroids and crop rows (Figure 2a). Plant centroids farther away from row vectors were eliminated to filter weeds. When delayed emergence were observed (in 2020), locations of newly emerged (n) plant centroids were appended if they were close to the row vectors (± 0.05 m) and did not overlap with older (n – 1) centroids (Figure 2b).

To accurately reconstruct surface elevation profiles, 80,000 random ground surface points (obtained from inverted canopy masks) were sampled from DEMs and linearly interpolated. Above ground canopy elevation was calculated by subtracting the surface elevation from the DEM. Individual plant height was estimated as the maximum canopy height within 0.03 m of plant centroids for developing canopies and 0.05 m for closed canopies, then corrected using the height bias measured from elevated control points. Plant count, PPD, and PH were evaluated using mean ($\bar{x}$), relative standard deviation (RSD), Pearson correlation (r), and root mean squared error (RMSE) (equations 5 to 7). Precision and recall (equations 8 and 9) were used to assess locational accuracy of plant centroids, where true positives (TP) indicated plants detected in the correct location, false negatives (FN) represented missed plants, and false positives (FP) represented detections that did not correspond to actual plants.

$$RSD = \frac{\sigma}{\bar{x}} \times 100 \quad (5)$$

$$r = \frac{\sum_{i=1}^{n}(x_i - \bar{x})(\hat{x}_i - \bar{\hat{x}})}{\sqrt{\sum_{i=1}^{n}(x_i - \bar{x})^2}\sqrt{\sum_{i=1}^{n}\left(\hat{x}_i - \bar{\hat{x}}\right)^2}} \quad (6)$$

$$RMSE = \sqrt{\frac{1}{n}\sum_{i=1}^{n}(\hat{x}_i - x_i)^2} \quad (7)$$

$$Precision = \frac{TP}{TP + FP} \quad (8)$$

$$Recall = \frac{TP}{TP + FN} \tag{9}$$

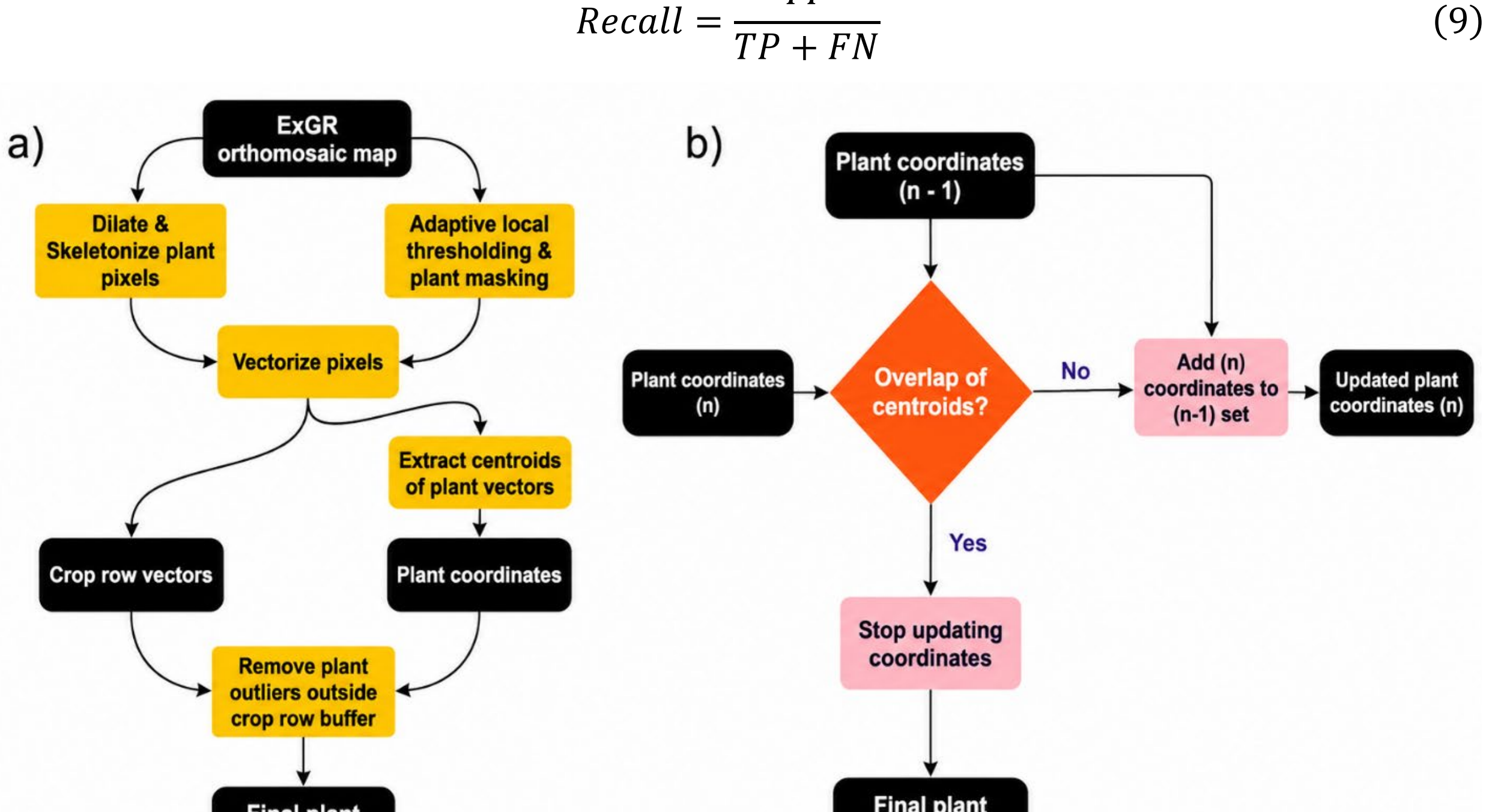

**Figure 2 Flowcharts representing (a) general plant count estimation from ExGR orthomosiac obtained after seedling emergence, and (b) multi-stage iterative process for updating plant count and location during delayed seedling emergence.**

## 2.4 Statistical modeling of plant growth parameters

Simple regression models interpolated linear or quadratic relationships between individual VIs and biophysical parameters (DBW, PNU, and PNC). Multiple linear regressions (MLRs) combined individual VIs with estimated PH and FCC to predict DBW, PNU, and PNC. Two data-splitting strategies – leave-one-year-out (LOYO) and treatment-hold-out (THO) – were implemented to reduce training bias by blocking spatiotemporal patterns present in the data. In LOYO, all observations from 2020 were withheld for independent validation of model generalizability across growing seasons. In THO, 25% of the unique year-treatment groups were reserved for independent validation, ensuring all observations from the same treatment within a given year were assigned to either training or validation. Also, THO train-test partitioning was repeated across ten seeding rates, and the models were evaluated using averages of mean absolute percentage error (MAPE) (equation 10) and coefficient of determination ($R^2$) metrics (equation 11).

$$MAPE = \frac{1}{n}\sum_{i=1}^{n}\left|\frac{\hat{y}_i - y_i}{y_i}\right| \times 100 \tag{10}$$

$$R^2 = 1 - \frac{\sum_{i=1}^{n}(\hat{y}_i - y_i)^2}{\sum_{i=1}^{n}(y_i - \bar{y}_i)^2} \tag{11}$$

2.5 Decision-tree models: feature selection and nested model development

Two ensemble decision-tree models – random forest regression (RFR) and extreme gradient boost (XGB) regression – were trained with spectral reflectance ($\rho_{mixed}$), PH, and FCC to predict DBW, PNU, and PNC. The RFR models used bagging approach, where the trees were trained in parallel and individual outcomes were aggregated for final decision. The XGB models learned by boosting, where trees were built sequentially and the outcome of one tree influenced the creation and training of subsequent trees. Both LOYO and THO validation strategies were implemented here, and THO hyperparameter tuning and feature selection were repeated across ten randomly grouped train-test splits to reduce training dependence on any single partition.

For RFR, randomized search was used for tuning number of trees, maximum tree depth, minimum samples for node splitting, minimum samples per leaf, number of features at each node split. For XGB, grouped (by year-treatments) cross-validation was used for selecting number of boosting iterations, with fixed tree depth, learning rate, subsampling fraction, and column sampling fraction. Feature importance was estimated with permutation importance for RFR and SHAP values for XGB. In THO, normalized feature importance scores were averaged across ten grouped train-test splits. Nested models were constructed by sequentially adding features from highest to lowest ranks and evaluated using $R^2$ and MAPE. The optimal set of features required for optimal model training was determined by a composite score (CS) that combined MAPE rank ($R_{MAPE}$), $R^2$ rank ($R_{R^2}$), and model parsimony rank ($R_F$), and the best feature set had the lowest CS value (equations 12 and 13). Hyperparameters were re-tuned for the optimal feature set, and model performance was evaluated on the independent validation data.

$$CS = 0.35\ R_{MAPE} + 0.40\ R_{R^2} + 0.25\ R_F \quad (12)$$

$$f_{best} = argmin(CS) \quad (13)$$

2.6 Nitrogen nutrition index for nitrogen stress detection

$N_c$ and NNI were derived from model estimated DBW and PNC, and binary categorization was performed, where plots with NNI < 1 were labeled as requiring fertilizer application (F) and those with NNI ≥ 1 were marked as not requiring additional fertilizer (NF). A multi-level stress categorization was also done to label plots as deficient (D) for NNI < 0.80, low (L) for 0.80 ≤ NNI < 0.98, sufficient (S) for 0.98 ≤ NNI < 1.08, and excess (E) for NNI ≥ 1.08 to grade and quantify N requirements for in-season PNM. The binary and multi-level labels were evaluated using class-specific precision and recall, and F1 score averaged across n classes (equation 14).

$$F1 = \sum_{i=1}^{n} \frac{2}{n} * \frac{Precision_i * Recall_i}{Precision_i + Recall_i} \quad (14)$$

## 3. Results

3.1 Critical nitrogen dilution curve and nitrogen nutrition index

The $N_c$ curve ($N_c = 2.45 * DBW_{act}^{-0.18}$) was established for certain high-yielding, medium-to-tall cotton varieties grown in the Texas Coastal Plains from biomass samples collected across three seasons (Figure 3). The treatment-level NNI distribution at different stages (Figure 4) showed that NNI values were mostly at sufficiency levels for all N treatments at initial vegetative growth (St0) and differential response to the treatments became evident in later stages.

T5 and T6 plots, subjected to optimal and higher N rates, respectively, had consistently higher NNI than other plots. Also, T4 plots subjected to early low N rates had NNI similar to T3 treated with optimal N at squaring, implying that fertilizer timing affects N uptake and status. Also, N was insufficient at St4 that had data from a single season (2022), compared to higher N availability at St5 that included 2020 and 2021 data, indicating possible early onset of maturity in 2022 due to high temperatures and zero rainfall (Table 2).

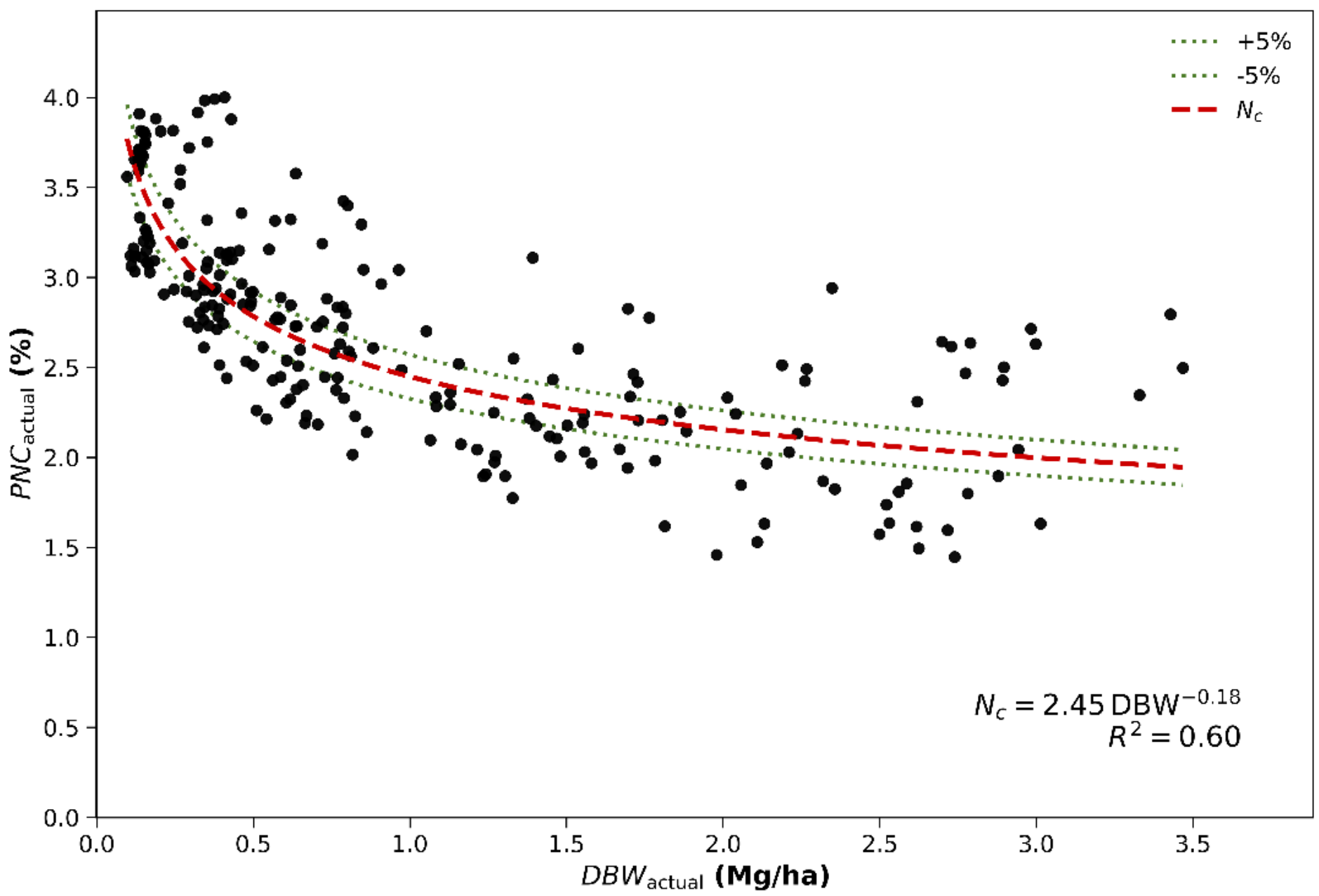

**Figure 3 Critical nitrogen concentration (Nc, red dashed line), within ±5% confidence interval (green dotted lines), interpolated from actual plant nitrogen concentration ($PNC_{actual}$, %) and dry biomass weight ($DBW_{actual}$, Mg/ha) obtained from biomass samples collected across three growing seasons.**

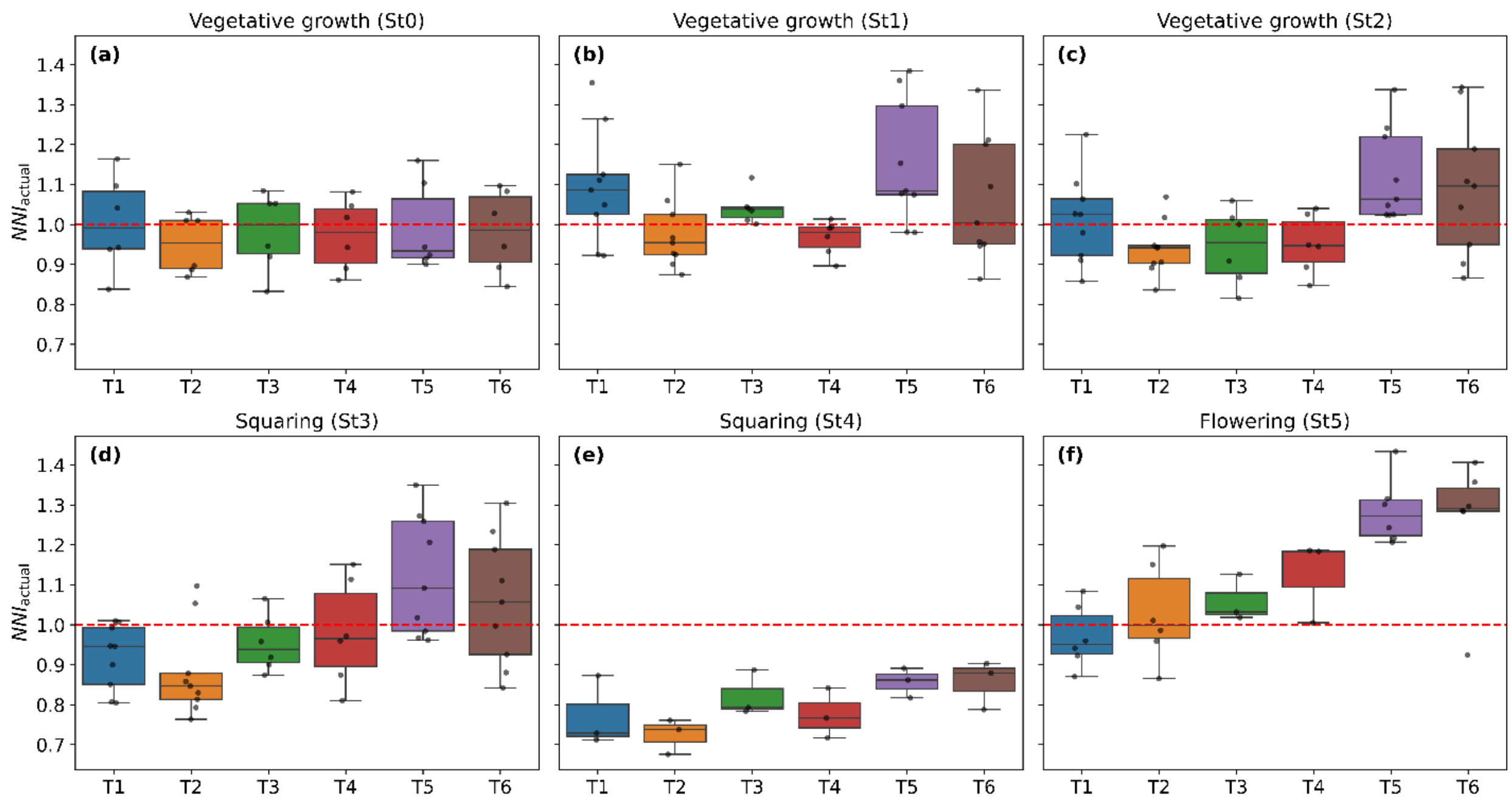

**Figure 4 The distribution of plot-level nitrogen nutrition index (NNI) for the different nitrogen treatments (T1, T2, T3, T4, T5, and T6) during vegetative growth (St0 to St2), squaring (St3 and St4), and flowering (St5). The black circles indicate the number of data points and red dotted line indicates nitrogen sufficiency (NNI = 1).**

### 3.2 Plant count and population density estimation

Plants were identified in their actual locations with good (> 0.80) to high (> 0.90) precision, highlighting the effectiveness of outlier pruning and delayed emergence identification methods (Table 4). The recall scores were good (> 0.80) in 2020 and 2021, however only average (0.65) in 2022, indicating some plants were not identified due to canopy clustering. Consequently, the difference between actual and estimated PPD was also high in 2022, compared to other years. Further, 2020 trial's estimated plot-level plant counts ($\hat{n}_{\text{plant}}$) were highly correlated (r = 0.96) with ground-truth ($n_{\text{plant}}$) and had low estimation error (RMSE = 29.67 plants/plot) (Figure 5). Two plots in the rear-end of the field with smaller dimensions (Figure 1, Rep 3, T2 and T5) and two more with uneven seedling emergence (Figure 1, Rep 1, T5 and T6) contributed to the low plant count (< 400 plants/plot). While $\widehat{PPD}$ for the smaller plots (3.98 and 4.43 plants/m$^2$) were closer to the plot-level average (4.44 plants/m$^2$), $\widehat{PPD}$ for plots with uneven seedling emergence (3.28 and 3.39 plants/m$^2$) were much lower, resulting in higher plot-level RSD in 2020.

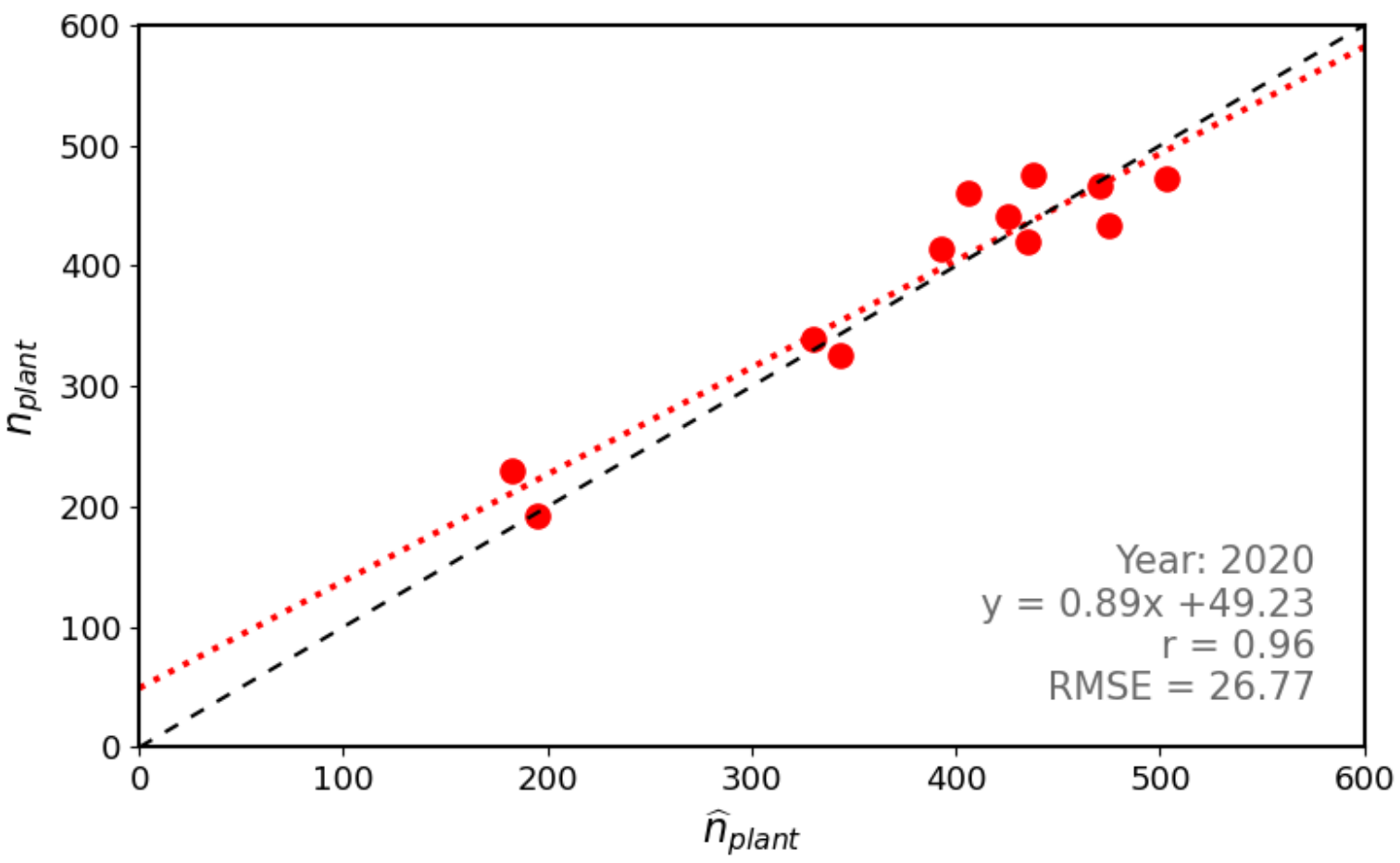

**Figure 5 Comparing actual plot-level plant counts ($n_{plant}$) with estimated plant counts ($\hat{n}_{plant}$) in the 2020 nitrogen management trial where cotton seedling emergence patterns were not uniform.**

**Table 4 Accuracy assessment of plant location and plant population density (PPD) estimation using precision-recall scores and manual ground truthing. The plot level variability in PPD was analyzed using mean and relative standard deviation (RSD).**

| Year | 1-m$^2$ sample analysis | | | | Plot-level $\widehat{PPD}$ (plants/m$^2$) | |
|---|---|---|---|---|---|---|
| | Precision | Recall | PPD (plants/m$^2$) | $\widehat{PPD}$ (plants/m$^2$) | Mean | RSD (%) |
| **2020** | 0.88 | 0.88 | 4.83 | 4.82 | 4.44 | 13.96 |
| **2021** | 0.96 | 0.83 | 8.00 | 6.96 | 6.58 | 4.86 |
| **2022** | 0.89 | 0.65 | 7.56 | 5.54 | 5.31 | 8.09 |

## 3.3 Above-ground plant height and fractional canopy cover estimation

Estimated plot-level plant height ($\widehat{PH}$) averages were highly correlated (r > 0.95) with ground-truth averages, with low RMSE (< 0.10 m) (Figure 6b and 6c) for 2021 and 2022. Although 2020's $\widehat{PH}$ had good correlation (r = 0.89), the estimation error was high (RMSE = 0.18 m), and the regression slope (0.63) was far from unity (Figure 6a). The weekly trends in plant height estimation (Figure 6d) showed a decrease in RMSE and increase in r as canopy cover attained full closure. At all stages, 2020 $\widehat{PH}$ had higher RMSE and lower r than 2021 and 2022 $\widehat{\text{PH}}$, indicating poor 3D canopy reconstruction arising from a relatively low lateral image overlap.

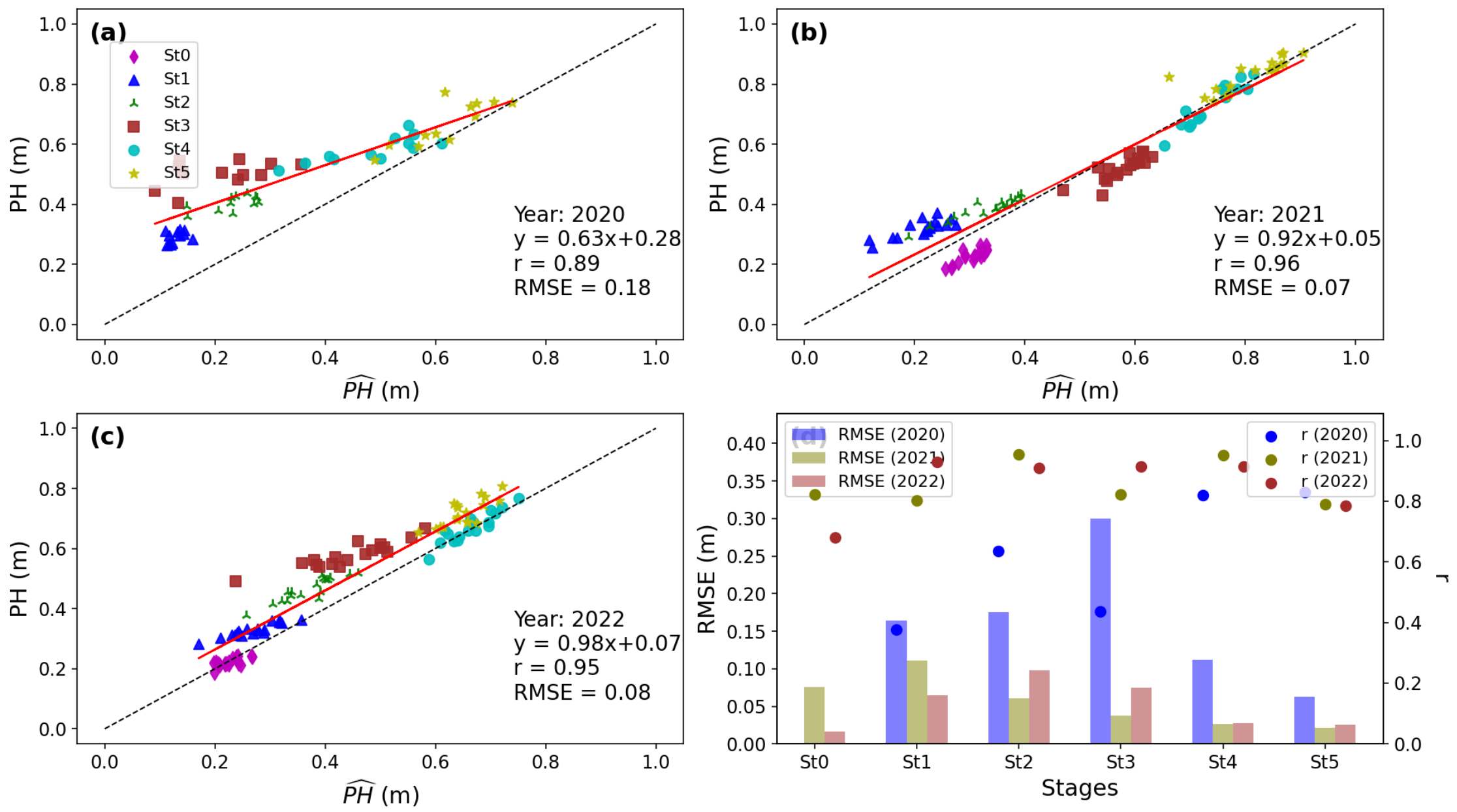

**Figure 6 Comparison of estimated plot-averaged above ground cotton plant height ($\widehat{PH}$, m) with actual plant height measurement (PH) averages for years 2020 (a), 2021 (b), and 2022 (c), and trends in RMSE (m) (bars) and Pearson correlation coefficient (r) (dots) from stages (St) 0 to 5 (d).**

### 3.4 Vegetation indices for estimating dry biomass weight, plant N uptake and concentration

VIs from calibrated multispectral images captured early-season variabilities in DBW ($0.82 \leq R^2 \leq 0.85$) and PNU ($0.77 \leq R^2 \leq 0.80$) consistently across all years (Figure 7). The relationship between the VIs and PNC demonstrated a clear seasonal pattern, which resulted in low $R^2$ (0.43 to 0.53). For simple regression, $\text{GNDVI}_{\text{canopy}}$ produced least testing error ($R^2 > 0.80$; MAPE = 31.28 %) for $\widehat{\text{DBW}}_{THO}$, followed closely by $\text{NDVI}_{\text{canopy}}$ and $\text{NDRE}_{\text{canopy}}$ (Table 5). $\text{NDVI}_{\text{canopy}}$ was also the only VI to predict $\widehat{\text{PNC}}_{THO}$ with test $R^2 > 0.50$ and MAPE = 12.11 %. For $\widehat{\text{PNU}}_{THO}$, $\text{CI}_{\text{rededge_mixed}}$ had the highest test $R^2 = 0.78$ and lowest MAPE = 29.38 %, followed closely by other $\text{VI}_{\text{mixed}}$ models. For $\widehat{\text{DBW}}_{LOYO}$, $\text{GNDVI}_{\text{mixed}}$, $\text{GNDVI}_{\text{canopy}}$, and $\text{NDVI}_{\text{canopy}}$ produced the least test errors (MAPE ≈ 30%) with $0.64 \leq R^2 \leq 0.70$. While $\text{CI}_{\text{rededge}}$ and $\text{NDVI}_{\text{canopy}}$ were still the best performing models for $\widehat{\text{PNU}}_{LOYO}$ and $\widehat{\text{PNC}}_{LOYO}$, their performances were significantly lower than the corresponding THO models. This highlighted the inability of the simple VI models to reliably predict unseen data, especially when seasonal patterns are present, as in $\widehat{\mathrm{PNC}}_{LOYO}$. In general, $\mathrm{VI}_{\mathrm{canopy}}$ were relatively better than $\mathrm{VI}_{\mathrm{mixed}}$ for $\widehat{\mathrm{DBW}}$ and $\widehat{\mathrm{PNC}}$, while $\mathrm{VI}_{\mathrm{mixed}}$ models were better for $\widehat{\mathrm{PNU}}$. Combining VIs with physical attributes using MLR marginally increased test MAPE for $\widehat{\mathrm{DBW}}_{THO}$, while MLR-based $\widehat{\mathrm{PNU}}_{THO}$ and $\widehat{\mathrm{PNC}}_{THO}$ were comparable to the best performing simple regression model (Table 6). Although test MAPE increased for MLR-based $\widehat{\mathrm{DBW}}_{LOYO}$, the best MLR models for $\widehat{\mathrm{PNU}}_{LOYO}$ and $\widehat{\mathrm{PNC}}_{LOYO}$ had significantly higher $R^2$, while MAPE remained similar to the best performing simple VI-based LOYO models. This demonstrates the potential for improving predictions of seasonally variable parameters by combining spectral and morphological features.

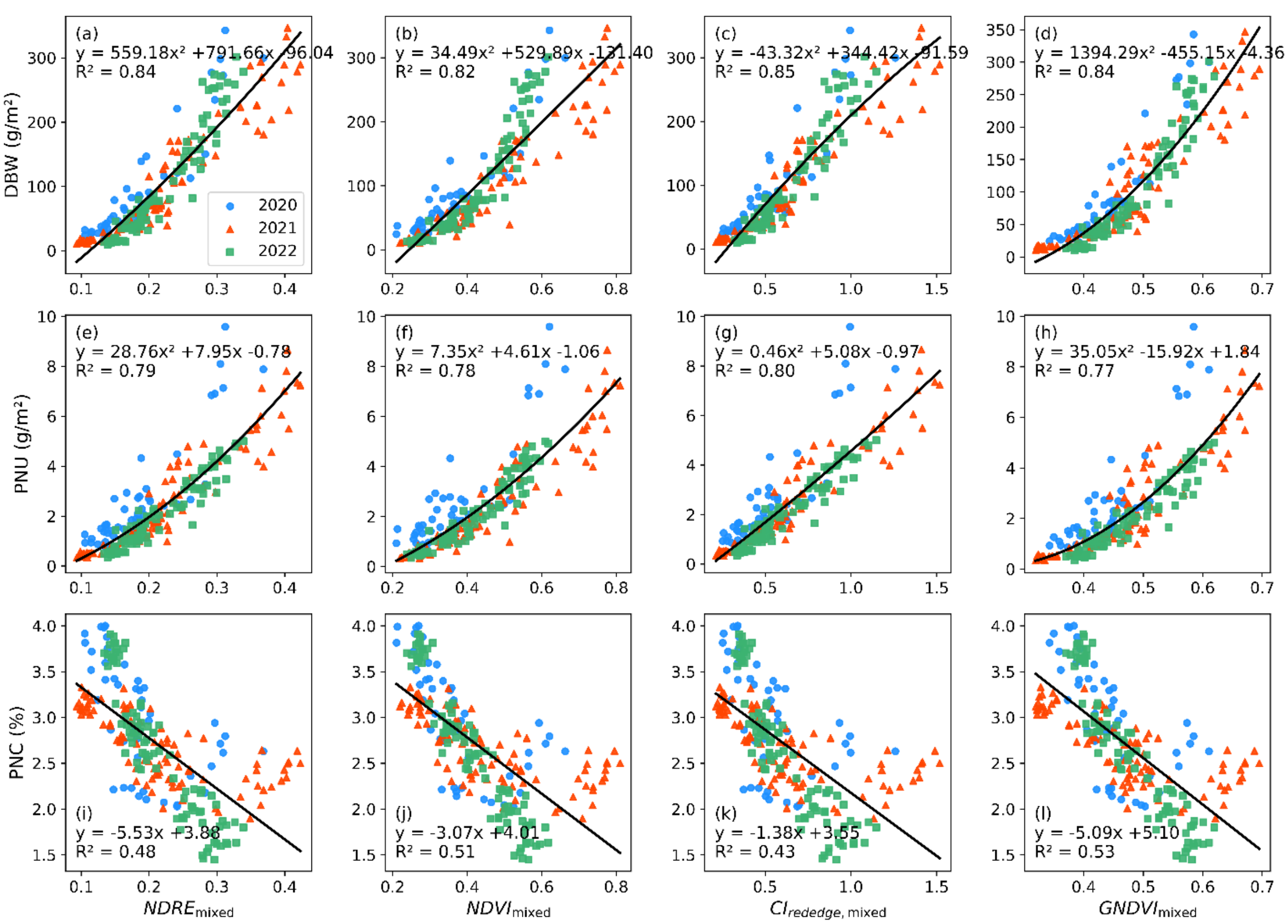

**Figure 7 Relationship between soil and canopy mixed vegetation indices (VImixed) and early-season cotton plant dry biomass weight, DBW (g/m$^2$) (a – d), plant N uptake, PNU (g/m$^2$) (e – h), and plant N concentration, PNC (%) (i – l).**

**Table 5 Training and Testing performance of trial-held-out (THO) and leave-one-year-out (LOYO) simple regression models for predicting cotton dry biomass weight (DBW, g/m$^2$), plant nitrogen uptake (PNU, g/m$^2$) and plant nitrogen concentration (PNC, %). Best performing models in each category are highlighted in bold.**

| **Model inputs** | | **Dry biomass weight (DBW, g/m$^2$)** | | | | **Plant nitrogen uptake (PNU, g/m$^2$)** | | | | **Plant nitrogen concentration (PNC, %)** | | | |
|---|---|---|---|---|---|---|---|---|---|---|---|---|---|
| | | **Train** | | **Test** | | **Train** | | **Test** | | **Train** | | **Test** | |
| | | **$R^2$** | **MAPE** | **$R^2$** | **MAPE** | **$R^2$** | **MAPE** | **$R^2$** | **MAPE** | **$R^2$** | **MAPE** | **$R^2$** | **MAPE** |
| Trial-Held-Out (THO) Regressions | | | | | | | | | | | | | |
| $VI_{mixed}$ | NDRE | 0.85 | 48.09 | 0.81 | 48.84 | 0.80 | 30.83 | 0.77 | 30.64 | 0.49 | 14.08 | 0.26 | 14.62 |
| | NDVI | 0.82 | 39.08 | 0.78 | 44.01 | 0.79 | 28.86 | 0.75 | 32.74 | 0.52 | 13.72 | 0.25 | 14.64 |
| | $CI_{rededge}$ | 0.86 | 46.19 | 0.82 | 48.05 | **0.81** | **29.58** | **0.78** | **29.38** | 0.45 | 14.62 | 0.20 | 15.19 |
| | GNDVI | 0.85 | 43.70 | 0.81 | 44.61 | 0.77 | 30.70 | 0.74 | 30.90 | 0.54 | 13.26 | 0.31 | 13.96 |
| $VI_{canopy}$ | NDRE | 0.81 | 35.60 | 0.79 | 34.35 | 0.72 | 30.47 | 0.67 | 29.97 | 0.49 | 14.02 | 0.32 | 14.56 |
| | NDVI | 0.85 | 33.17 | 0.82 | 34.31 | 0.75 | 28.48 | 0.71 | 32.15 | **0.64** | **11.91** | **0.52** | **12.11** |
| | $CI_{rededge}$ | 0.77 | 43.66 | 0.75 | 42.86 | 0.69 | 33.18 | 0.64 | 32.76 | 0.46 | 14.41 | 0.30 | 14.83 |
| | GNDVI | **0.83** | **29.94** | **0.81** | **31.28** | 0.72 | 27.17 | 0.67 | 28.36 | 0.55 | 13.06 | 0.36 | 13.70 |
| Leave-One-Year-Out (LOYO) Regressions | | | | | | | | | | | | | |
| $VI_{mixed}$ | NDRE | 0.87 | 47.77 | 0.71 | 39.11 | 0.90 | 25.44 | 0.46 | 38.09 | 0.50 | 13.33 | 0.23 | 15.83 |
| | NDVI | 0.84 | 42.99 | 0.69 | 44.15 | 0.89 | 21.29 | 0.42 | 40.36 | 0.51 | 13.57 | 0.34 | 14.46 |
| | $CI_{rededge}$ | 0.88 | 45.95 | 0.72 | 40.99 | **0.90** | **24.47** | **0.49** | **35.45** | 0.46 | 13.70 | 0.13 | 16.88 |
| | GNDVI | **0.87** | **43.18** | **0.70** | **30.38** | 0.87 | 23.96 | 0.43 | 33.79 | 0.55 | 12.51 | 0.28 | 15.36 |
| $VI_{canopy}$ | NDRE | 0.86 | 32.31 | 0.60 | 38.33 | 0.81 | 25.56 | 0.44 | 33.37 | 0.57 | 12.43 | -0.01 | 18.24 |
| | NDVI | 0.88 | 33.69 | 0.66 | 31.50 | 0.86 | 24.11 | 0.40 | 33.62 | **0.67** | **11.21** | **0.37** | **13.65** |
| | $CI_{rededge}$ | 0.85 | 39.36 | 0.42 | 47.61 | 0.80 | 27.88 | 0.36 | 38.45 | 0.56 | 12.57 | -0.16 | 19.42 |
| | GNDVI | 0.87 | 28.93 | 0.64 | 28.77 | 0.80 | 24.24 | 0.44 | 26.76 | 0.62 | 11.49 | 0.02 | 17.30 |

**Table 6 Training and Testing performance of best performing multiple linear regression (MLR) and decision-tree – random forest regression (RFR) and extreme gradient boost regression (XGB) – models that combined spectral (vegetation indices or reflectance) and morphological features to predict cotton dry biomass weight (DBW, g/m²), plant nitrogen uptake (PNU, g/m²) and plant nitrogen concentration (PNC, %).**

| Model inputs | Dry biomass weight (DBW, g/m²) | | | | Plant nitrogen uptake (PNU, g/m²) | | | | Plant nitrogen concentration (PNC, %) | | | |
|---|---|---|---|---|---|---|---|---|---|---|---|---|
| | Train | | Test | | Train | | Test | | Train | | Test | |
| | $R^2$ | MAPE | $R^2$ | MAPE | $R^2$ | MAPE | $R^2$ | MAPE | $R^2$ | MAPE | $R^2$ | MAPE |
| Trial-Held-Out (THO) Regressions | | | | | | | | | | | | |
| MLR1 | $f_{MLR}(CI_{rededge_mixed}, PH)$ | | | | $f_{MLR}(CI_{rededge_mixed}, FCC)$ | | | | $f_{MLR}(NDVI_{canopy}, PH, FCC)$ | | | |
| | 0.88 | 37.45 | 0.84 | 39.99 | 0.84 | 26.34 | 0.80 | 31.23 | 0.67 | 11.03 | 0.56 | 11.27 |
| MLR2 | $f_{MLR}(CI_{rededge_mixed}, PH, FCC)$ | | | | $f_{MLR}(CI_{rededge_mixed}, PH, FCC)$ | | | | $f_{MLR}(NDVI_{canopy}, FCC)$ | | | |
| | 0.88 | 38.75 | 0.84 | 41.19 | 0.84 | 27.09 | 0.80 | 33.12 | 0.66 | 11.04 | 0.56 | 11.45 |
| MLR3 | $f_{MLR}(GNDVI_{mixed}, PH, FCC)$ | | | | $f_{MLR}(CI_{rededge_canopy}, PH, FCC)$ | | | | $f_{MLR}(NDVI_{canopy}, PH)$ | | | |
| | 0.88 | 38.95 | 0.84 | 42.68 | 0.84 | 27.81 | 0.80 | 34.85 | 0.64 | 11.90 | 0.51 | 12.19 |
| RFR | $f_{RFR}(PH, NIR, FCC)$ | | | | $f_{RFR}(PH, FCC, NIR)$ | | | | $f_{RFR}(FCC, NIR, PH, B, R, RE)$ | | | |
| | 0.99 | 8.97 | 0.88 | 23.14 | 0.93 | 19.12 | 0.84 | 20.61 | 0.95 | 4.11 | 0.85 | 7.82 |
| XGB | $f_{XGB}(PH, NIR, FCC, B, RE, G)$ | | | | $f_{XGB}(PH, FCC, NIR)$ | | | | $f_{XGB}(FCC, PH, NIR, R, RE, B)$ | | | |
| | 0.99 | 6.08 | 0.87 | 21.91 | 0.98 | 12.07 | 0.81 | 21.40 | 0.99 | 1.99 | 0.86 | 7.66 |
| Leave-One-Year-Out (LOYO) Regressions | | | | | | | | | | | | |
| MLR1 | $f_{MLR}(CI_{rededge_mixed}, PH)$ | | | | $f_{MLR}(GNDVI_{canopy}, FCC)$ | | | | $f_{MLR}(NDVI_{canopy}, PH, FCC)$ | | | |
| | 0.90 | 28.08 | 0.72 | 58.37 | 0.87 | 32.71 | 0.62 | 35.67 | 0.69 | 10.45 | 0.42 | 12.94 |
| MLR2 | $f_{MLR}(GNDVI_{mixed}, PH)$ | | | | $f_{MLR}(NDRE_{canopy}, FCC)$ | | | | $f_{MLR}(NDVI_{canopy}, FCC)$ | | | |
| | 0.91 | 29.10 | 0.68 | 67.90 | 0.88 | 31.84 | 0.61 | 39.98 | 0.69 | 10.56 | 0.38 | 13.25 |
| MLR3 | $f_{MLR}(NDRE_{mixed}, PH)$ | | | | $f_{MLR}(GNDVI_{mixed}, FCC)$ | | | | $f_{MLR}(NDVI_{canopy}, PH)$ | | | |
| | 0.91 | 29.47 | 0.68 | 67.28 | 0.88 | 28.91 | 0.57 | 36.57 | 0.67 | 11.12 | 0.34 | 13.83 |
| RFR | $f_{RFR}(PH, NIR, FCC)$ | | | | $f_{RFR}(PH, FCC, B, NIR, G)$ | | | | $f_{RFR}(FCC, B)$ | | | |
| | 0.98 | 11.71 | 0.84 | 24.64 | 0.98 | 9.09 | 0.68 | 31.01 | 0.86 | 6.78 | 0.41 | 13.37 |
| XGB | $f_{XGB}(PH, NIR, FCC)$ | | | | $f_{XGB}(PH, FCC, B, NIR, G, R)$ | | | | $f_{XGB}(FCC, B)$ | | | |
| | 0.99 | 11.87 | 0.84 | 25.27 | 0.99 | 8.80 | 0.73 | 30.14 | 0.86 | 7.12 | 0.32 | 13.98 |

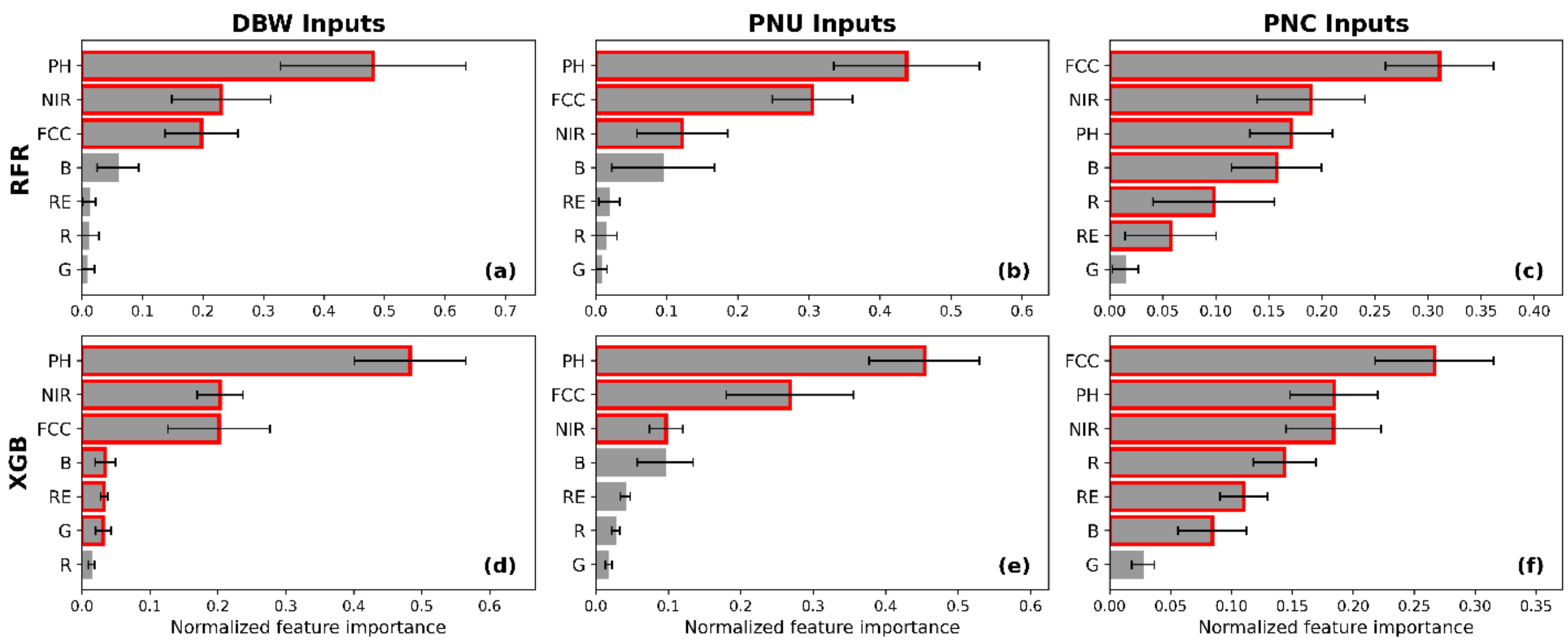

**Figure 8 Normalized feature importance ranking based on permutation importance for random forest regression (RFR) models (a – c) and shapely additive explanations (SHAP) for extreme gradient boost (XGB) models (d – f), in predicting dry biomass weight, DBW (g/m$^2$) (a, d), plant N uptake, PNU (g/m$^2$) (b, e), and plant N concentration, PNC (%) (c, f). Standard deviation bars represent variation from ten random trial-held-out (THO) data splits and red highlights represent optimal features used for final training.**

### 3.5 Feature selection and decision tree models for plant growth parameter estimation

Feature rankings for RFR and XGB showed that $\widehat{\text{PH}}$ was most important for $\widehat{\text{DBW}}_{THO}$ and $\widehat{\text{PNU}}_{THO}$, and $\widehat{\text{FCC}}$ for $\widehat{\text{PNC}}_{THO}$, highlighting the significance of morphological features in predicting biophysical plant parameter (Figure 8). Both RFR and XGB exhibited parsimony by eliminating one or more features, with $\widehat{\text{PNU}}_{THO}$ requiring the least number of features (highlighted in red, Figure 8). For THO, the performance of RFR and XGB models were very similar, and both showed significant improvement in $R^2$ and reduction in MAPE compared to the VI-based models for all parameters (Table 6). While inclusion of morphology did not improve $\widehat{\text{DBW}}_{THO}$ through MLR, the test MAPE decreased by more than 7% with decision-trees compared to simple VI regression. There was also a notable increase in test $R^2$ ($\geq 0.70$) for $\widehat{\text{PNC}}_{THO}$ predicted by the decision-tree models. Likewise, $\widehat{\text{DBW}}_{LOYO}$ and $\widehat{\text{PNU}}_{LOYO}$ predictions also improved for the decision-tree models compared to simple regression and MLR. However, for $\widehat{\text{PNC}}_{LOYO}$, XGB performed worse than the best simple regression model, and RFR performed only on par with the best MLR model. This indicated that the decision-tree models generalized well for seasonally invariable $\widehat{\mathrm{DBW}}_{LOYO}$ and $\widehat{\mathrm{PNU}}_{LOYO}$, but were less effective for PNC, where strong interannual variability reduced prediction accuracy when an entire year was withheld.

### 3.6 Plant stress level classification and fertilizer management decision with nitrogen nutrition index

$N_c$ dilution coefficients derived from the THO and LOYO training outcomes of the RFR and XGB models were similar to the actual $N_c$ curve coefficients (Figure 9). Although the relationship between $\widehat{\mathrm{DBW}}_{LOYO}$ and $\widehat{\mathrm{PNC}}_{LOYO}$ produced stronger $N_c$ fits than their corresponding THO predictions, $\widehat{\mathrm{NNI}}_{LOYO}$ had poor association with $NNI_{actual}$, likely due to the high uncertainty in $\widehat{\mathrm{PNC}}_{LOYO}$ for both models (Table 6). This reiterated the inability of LOYO decision-tree models to generalize when seasonal patterns dominate. In contrast, $\widehat{\mathrm{NNI}}_{THO}$ had high training $R^2 > 0.85$ and low errors (RMSE < 0.06; MAPE < 4.50%) with average testing $R^2$ (≈ 0.53) and low errors (RMSE ≈ 0.09; MAPE ≈ 7.5%) (Figure 9). Consequently, THO validation showed good binary classification performance, with the XGB (F1 = 0.75) marginally outperforming RFR (F1 = 0.70) (Table 7). Also, XGB outputs identified N-stressed plots (F) with higher precision and recall, which is important for PNM decisions because failing to detect stress is generally more detrimental for crop growth and yield than misclassifying healthy plots (NF) as being stressed. Again, XGB (F1 = 0.59) performed better that RFR (F1 = 0.52) in categorizing NNI-based stress levels (Table 8). However, the evident class imbalance with only four plots representing N deficiency (D), made it difficult to justify whether the XGB outputs were better at differentiating D plots from low N (L) plots. The overall LOYO binary N-input categorization outcomes from both RFR (F1 = 0.61) and XGB (F1 = 0.58) were lower than the corresponding THO scores and had < 50% recall of F plots. LOYO outcomes for multi-level stress categorization were also significantly poor (RFR: F1 = 0.35; XGB: F1 = 0.27).

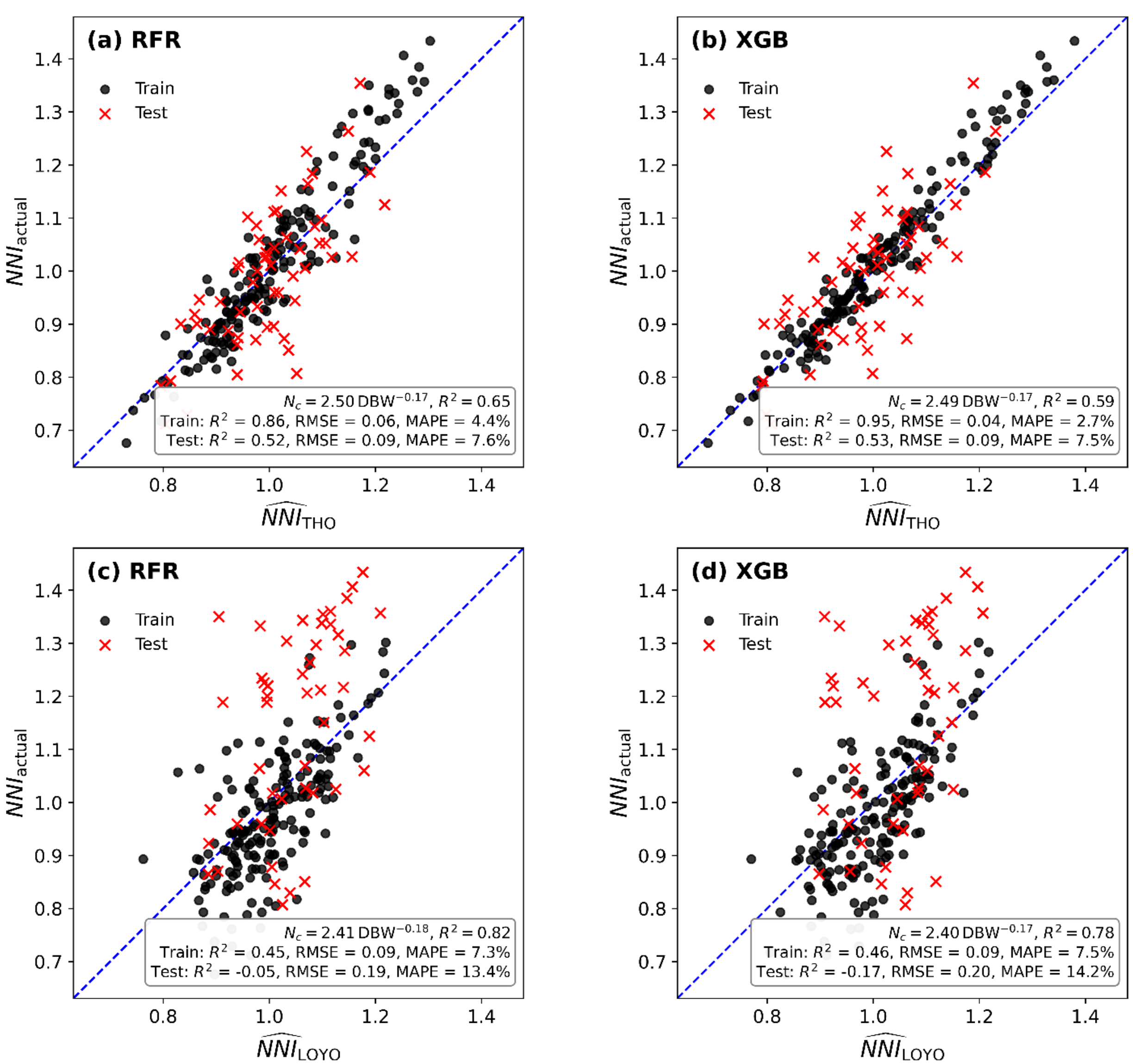

**Figure 9 Comparison of actual and predicted nitrogen nutrition index (NNI) derived from dry biomass weight (DBW) and plant nitrogen concentration (PNC) estimated by trial-held-out (THO) random forest regression (RFR) (a) and THO extreme boost regression (XGB) (b), leave-one-year-out (LOYO) RFR (c), and LOYO XGB (d).**

**Table 7 Trial-held-out (THO) validation of nitrogen fertilizer management decisions using nitrogen nutrition index ($\widehat{\mathrm{NNI}}_{THO}$) from random forest regression (RFR) and extreme gradient boost regression (XGB) models. F: Fertilizer application if NNI < 1, and NF: No Fertilizer application if NNI ≥ 1. The average F1 score of the categorization is highlighted in bold.**

| Fertilizer Decision | Random Forest Regression | | | Extreme Gradient Boost Regression | | |
|---|---|---|---|---|---|---|
| | F | NF | Recall | F | NF | Recall |
| **F** | 19 | 8 | 0.71 | 21 | 6 | 0.77 |
| **NF** | 9 | 21 | 0.70 | 8 | 22 | 0.73 |
| **Precision** | 0.68 | 0.72 | **0.70** | 0.72 | 0.79 | **0.75** |

**Table 8 Trial-held-out (THO) validation of nitrogen stress levels categorization – deficient (D), low (L), sufficient (S), excess (E) – using nitrogen nutrition index ($\widehat{\mathbf{NNI}}_{THO}$) from random forest regression (RFR) and extreme gradient boost regression (XGB) models. The average F1 score of the categorization is highlighted in bold.**

| N Status | Random Forest Regression | | | | | Extreme Gradient Boost Regression | | | | |
|---|---|---|---|---|---|---|---|---|---|---|
| | D | L | S | E | Recall | D | L | S | E | Recall |
| **D** | 1 | 3 | 0 | 0 | 0.25 | 3 | 1 | 0 | 0 | 0.75 |
| **L** | 0 | 14 | 8 | 0 | 0.64 | 1 | 14 | 6 | 1 | 0.63 |
| **S** | 0 | 3 | 10 | 4 | 0.59 | 0 | 4 | 9 | 4 | 0.53 |
| **E** | 0 | 2 | 5 | 7 | 0.50 | 0 | 2 | 6 | 6 | 0.43 |
| **Precision** | 1.00 | 0.64 | 0.43 | 0.64 | **0.52** | 0.75 | 0.67 | 0.43 | 0.55 | **0.59** |

# 4. Discussion

4.1 Remote sensing calibration and feature extraction protocols for experiment repeatability

Aerial remote sensing of large agricultural fields characterized by substantial spatial and temporal variabilities in plant growth and environment conditions require robust data collection and calibration protocols to ensure consistent measurements for modeling and decision-making. Image acquisition and calibration protocols, like the use of fixed-exposure settings (Swaminathan et al. 2024b), illumination correction with on-board DLS (Swaminathan et al. 2024a), and post-processing calibration with in-field targets minimized reflectance errors and produced consistent VIs across different growth stages and seasons (Figure 7). The multi-step plant counting method addressed management challenges related to uneven distribution of seeds and inconsistent germination. The plant count accuracy was comparable to several vision and deep-learning approaches used for cotton plant counting (Chen, R. et al. 2018; Feng et al. 2020; Jiang, Y. et al. 2019). However, reliable identification of individual plants within dense clusters with variable plant spacing remains a challenge. Deriving field terrain directly from DEMs, rather than relying on bare-ground digital terrain maps, reduced inconsistencies caused by differences in camera orientation, image overlap, and GPS accuracy across data collection events. While earlier studies extracted plant height as percentiles from DEM and 3D point clouds (Feng et al. 2018; Lu et al. 2021; Malambo et al. 2018), this study accurately determined individual plant height using plant locations, an approach particularly effective under highly variable growth patterns. Plant height estimation accuracy was low initially (St0 to St2) (Figure 6d) because smaller canopies that are farther away from the camera are more susceptible to mixed pixels that affect photogrammetric feature mapping and depth estimation. Also, reconstruction accuracy of fixed, structurally defined elevation targets does not necessarily reflect plant reconstruction accuracy, because plant canopies have irregular structures that move from wind and other disturbances. Although higher image overlaps and lower ground sampling distance can improve 3D reconstruction, they also increase battery consumption per area surveyed.

### 4.2 Scope for improving in-season cotton nitrogen status estimation models

Developing generalized prediction models that work across genotypes, environments, and management systems is challenging due to limited availability of comprehensive remote sensing and ground truth data. Most published N management studies performed random train-test splits, without explicitly accounting for spatial correlation among plots, temporal correlation across sampling dates, and seasonal variability that may limit model generalizability. Given similar data size limitations, THO and LOYO validations were performed to partially and completely offset spatiotemporal biases. Although VIs correlate well with several plant physiological processes, they are not entirely reliable for fertilizer management decisions because spectral signatures and plant growth characteristics vary considerably between varieties and phenotypes. Figure 7 showed that seasonal parameters like PNC were poorly estimated by VIs models especially in LOYO validation. By combining morphological and spectral features, models captured genotypic and phenotypic variability that may be overlooked when complex plant traits are represented by a single feature type. For example, integration of PH and FCC, which represent vertical and horizontal canopy expansion, respectively, with band reflectance indicating canopy health can help distinguish contrasting growth patterns, such as taller plants with limited canopy vigor and shorter plants with healthier canopies, thereby improving model applicability across diverse varieties and phenotypes. Feature integration significantly improved THO validation for all parameters and LOYO validation for parameters less affected by seasonal variability. The relatively better LOYO validation of DBW and PNU decision-tree models, and the contrasting poor LOYO results for seasonally variable PNC were consistent with observations from Thieme et al. (2025). Nonetheless, LOYO validation of MLR and decision-tree PNC outputs were still better than those of simple regression. The parsimonious decision-tree models achieved THO performance comparable to those trained with substantially larger and redundant input features (Chen, X. et al. 2023; Cui et al. 2025; Peng et al. 2024; Tian et al. 2024).

4.3 Remote sensing of critical nitrogen concentration and nitrogen nutrition index

Biomass samples collected across three growing seasons captured phenological variations in crop response to management and environment, establishing a representative ground-truth $N_c$ curve for similar cotton varieties grown in the Texas Coastal Plains. The estimated cotton $N_c$ coefficients differed from those reported in other studies (Hou et al. 2021; Qin et al. 2025; Wang et al. 2025), emphasizing the need to periodically recalculate $N_c$ for different cultivars and regions. Also, with proper data collection, image processing, and modeling techniques, $N_c$ for a new variety can be estimated from long-term remote sensing data. The THO validation performance of NNI (Figure 9) was comparable to studies that directly estimated cultivar and region-specific NNI using machine learning (Jia et al. 2025; Pei et al. 2023). However, estimating NNI from relatively robust biophysical parameters like DBW, PNU or PNC provides flexibility to independently recalibrate N requirements for new crops and fields, without retraining NNI models on new data.

### 4.4 Limitations and future work

The LOYO validation of the decision-tree models showed that additional data representing more diverse management, environment, and phenotypic response is needed to improve model generalizability. However, such large models also require extensive ground-truth measurements of biophysical parameters, which are time, cost, and labor intensive. The development of in-vivo plant ion and nutrient monitoring electrodes and their integration with remote sensing data demonstrate can be effective alternatives for destructive nutrient analysis (Bulacio Fischer et al. 2025; Zhai et al. 2025). Eventually, with the ease in availability of high-resolution spatiotemporal data, advanced models like long-short-term-memory (LSTM) can be trained with additional structural features like leaf area index, weather, and management inputs to improve estimation of seasonally variable parameters and track in-season crop development and critical fertilizer uptake phases (Moussaid et al. 2025). Lastly, the framework developed in this study only supports detection of nitrogen stress levels. Therefore, additional research is needed in developing data-driven decision support tools that translate N status indicators, such as NNI, into quantitative N fertilizer recommendations.

## 5. Conclusions

Monitoring cotton canopy growth and development is essential for early detection of nitrogen deficiency to help with timely N management decisions. UAV-based multispectral images provide high-resolution spatiotemporal data pertaining to canopy reflectance spectra and morphology. The findings of this study highlight the importance of robust calibration procedures and reliable feature extraction methods to generate spatially and temporally diverse datasets with minimal measurement uncertainty. Through feature importance ranking and nested model evaluation, this study demonstrated that combining spectral band reflectance and physical plant features improved predictions of DBW, PNU, and PNC. The improved performance was due to the complementary nature of the input features, where spectral reflectance represented crop health and morphological features represented canopy structure and development. When combined, these features enabled models to capture a broader range of variability associated with crop development and N uptake. Both RFR and XGB consistently produced reliable estimates of biophysical parameters like DBW and PNU that do not have evident seasonal patterns. They also estimated plant nitrogen concentration well when partially exposed to seasonal patterns using trial-held-out data splitting. The critical nitrogen ($N_c$) dilution curve was interpolated from estimated DBW and PNC and validated used ground-truth measurements NNI, derived from biophysical parameters predicted by XGB, was relatively more effective in identifying plots requiring fertilizer application and moderately effective in grading the stress levels. Since $N_c$ and NNI are highly variety and region specific, deriving NNI from stable crop growth parameter estimates, will facilitate easier integration of NNI-based decision support tools for wider varieties and environments. In conclusion, this study developed a generalizable framework for reliably estimating NNI and demonstrated the potential of UAV-based spectral and morphological plant features in developing physically meaningful machine learning models to support PNM for cotton production.

## Acknowledgements

The authors acknowledge the support provided by the members of the lab, notably Roy Graves, Chiranjibi Poudyal, Giordano Fontana, and Jeffrey Siegfried, and other field personnel at Texas A&M University and Texas A&M Agrilife Research who assisted with data collection, sample processing, and field management throughout the study. Their contributions were invaluable to the successful completion of this research.